\documentstyle[sprocl,epsfig]{article}

\def\be{\begin{equation}}
\def\ee{\end{equation}}
\def\bc{\begin{center}}
\def\ec{\end{center}}
\def\dst{\displaystyle\strut}
\def\ov{\over\displaystyle\strut}

\def\t0{\tau_0}

\def\bea{\begin{eqnarray}}
\def\eea{\end{eqnarray}}
\def\l({\left(}
\def\r){\right)}

\def\ket#1{\vert#1\rangle}
\def\bra#1{\langle#1\vert}
\def\brak#1#2{\langle#1\vert #2\rangle}

\def\bp{{\bf{p}}}
\def\bk{{\bf{k}}}
\def\bpi{{\bf{\pi}}}
\def\bxi{{\bf{\xi}}}

\def\apd{\hat a^{\dag} (\bp)}

\def\ri{\right)}
\def\lef{\left(}
\def\dst{\displaystyle\phantom{|}}
\def\ov{\over\dst}

\def\eps{\epsilon}
\def\bak{{\bf K}}
\def\dek{{\bf \Delta k}}

\begin{document}

\title{ON BOSE-CONDENSATION OF WAVE-PACKETS \\ IN HEAVY ION COLLISIONS}

\author{T. CS\"ORG\H O, \underline{J. ZIM\'ANYI} 
}

\address{MTA KFKI RMKI,\\ H-1525  Budapest 114, POB 49, Hungary \\
E-mails: jzimanyi@sunserv.kfki.hu, csorgo@sunserv.kfki.hu}

\maketitle\abstracts{ 
A recently obtained exact analytic solution to the wave-packet version
of the pion - laser model is presented. In the rare gas limit, a thermal
behaviour is found while the dense gas limiting case corresponds
to Bose-Einstein condensation of wave-packets to the wave-packet mode
with the smallest possible energy. 
}

\section{Introduction}

The study of the statistical properties of quantum systems has a long history
with important recent developments. In high energy physics quantum statistical
correlations are studied in order to infer the space-time dimensions of
the intermediate state formed in elementary particle reactions. In high
energy heavy ion collisions hundreds of bosons are created. The correct 
theoretical description of their correlations is difficult.
In this conference contribution we present a contradiction free treatement
 of multi-particle Bose-Einstein correlations for arbitrary large
number of particles.

\section{Model Assumptions:}

A model system is described as 
\be
\hat \rho \, = \, \sum_{n=0}^{\infty} \, {p}_n \, \hat \rho_n,
\ee
the density matrix of the whole system is normalized
to one. Here $ \hat \rho_n $ is the density matrix for 
events with fixed particle number $n$, which is normalized 
also to one. The probability for such an event is $ p_n $.

The multiplicity distribution is described by the set
$\left\{ {p}_n\right\}_{n=0}^{\infty}$.

The density matrix of a system with a fixed number 
of boson wave packets has the form 
\be
\hat \rho_n \, =\,  \int d\alpha_1 ... d\alpha_n
\,\,\rho_n(\alpha_1,...,\alpha_n)
\,\ket{\alpha_1,...,\alpha_n} \bra{\alpha_1,...,\alpha_n},
\ee
where $\ket{\alpha_1,...,\alpha_n}$ denote properly normalized
$n$-particle wave-packet boson states.

The wave packet creation operator is given as
\be
\alpha_i^{\dag}  \, = \, \int {d^3\bp \over
(\pi \sigma^2 )^{3\over 4} } \ 
	\mbox{\rm e}^{
		-{( \bp- \bpi_i)^{2}\over 2 \sigma_i^{2}}
	-  i \bxi_i (\bp - \bpi_i) + i \omega(\bp) ( t-t_i) 
	 } \ \apd .
\label{e:4}
\ee
The  commutator  
\be
\left[ \alpha_i, \alpha_j^{\dag} \right] \, = \,
	\brak{\alpha_i}{\alpha_j}
\ee
vanishes only in the case, when the wave packets do not overlap.
 
Here 
{$\alpha_i \, = \, (\bxi_i, \bpi_i, \sigma_i,t_i)$} refers 
to the center 
{in space, in momentum space, the width}
in momentum space and 
{the production time}, respectively. 

For simplicity we assume that
{ 
$\alpha_i \, =\, (\bpi_i, \bxi_i, \sigma, t_0)$.}

We call the attention to the fact that although one cannot 
attribute exactly defined values for space and momentum at the same
time, one can define precisely the $ \bpi_i, \bxi_i $ parameters.

The $n$ boson states, normalized to unity, are given as
\be
\ket{\ \alpha_1, \ ...\ , \ \alpha_n} \, = \,
 {1\over \sqrt{ \displaystyle{\strut
 \sum_{\sigma^{(n)} }
\prod_{i=1}^n
\brak{\alpha_i}{\alpha_{\sigma_i}}
 } } }
\  \alpha^{\dag}_n  \ ... \
\alpha_1^{\dag} \ket{0}.
\label{e:expec2}
\ee
Here $\sigma^{(n)}$ denotes the set of all the permutations of 
the indexes $\left\{1, 2, ..., n\right\}$ 
and the subscript  $_{\sigma_i}$ denotes the index that
replaces the index $_i$ in a given permutation from $\sigma^{(n)}$. 
The normalization factor contains a sum of $ n! $ term. These terms
contain $ n$ different $\alpha_i $ parameters. Thus the further calculation
with these normalized states seems to be extremely difficult.

\section{A New Type of Density Matrix \label{s:3}}

There is one special density matrix, however, for which one can
overcome the difficulty, related to the
non-vanishing overlap of many hundreds of wave-packets,
even in an explicit analytical manner. This density matrix is the 
product uncorrelated single particle matrices multiplied with 
a correlation factor, related to the induced emission: 
\be
\rho_n(\alpha_1,...,\alpha_n) \, = \, {\dst 1 \ov {\cal N}{(n)}}
\lef \prod_{i=1}^n \rho_1(\alpha_i) \ri \,
\lef\sum_{\sigma^{(n)}} \prod_{k=1}^n \, 
\brak{\alpha_k}{\alpha_{\sigma_k}}
\ri .
\label{e:dtrick}
\ee
Normalization  to  one  yields ${\cal N}(n)$.
 The weight factors describe 
{induced emission}:
\be
{\dst \rho_n(\alpha_1,...,\alpha_n) \ov \prod_{j=1}^n \rho_1(\alpha_j)}
 \, = \,{\dst    
\sum_{\sigma^{(n)}} \prod_{k=1}^n \, 
\brak{\alpha_k}{\alpha_{\sigma_k}}  
\ov {\cal N}{(n)} } \qquad \mbox{\rm
{ (overlap)}}.
 \ee
This is maximal if all are emitted with same
$ \alpha_1 $:
\be
{\dst \rho_n(\alpha_1,...,\alpha_1)  \ov [\rho_1(\alpha_1)]^n }
 \, = \, {\dst   n!   \ov {\cal N}{(n)} }, \qquad \mbox{\rm 
{(full overlap)}},
\ee
and minimal if the overlap is negligible,
\be
{\dst \rho_n(\alpha_1,...,\alpha_n) \ov \prod_{j=1}^n \rho_1(\alpha_j)}
 \, = \, {\dst   1   \ov {\cal N}{(n)} } \qquad 
	\mbox{\rm 
{(no overlap)}}.
 \ee
Thus we have wildly fluctuating weight. E.g. for 800 pions 
the induced emission weight fluctuates between 
{
$[1,800!] \simeq [1, 10^{1977}] $}.
With this special density matrix one can proceed with the 
calculations.

\subsection{ Single-Particle Density Matrix : }

For the sake of simplicity we assume a factorizable Gaussian form
for the distribution of the parameters of the single-particle 
states:
\bea
\rho_1(\alpha)& = &\rho_x(\bxi)\, \rho_p (\bpi)\, \delta(t-t_0), \cr
\rho_x(\bxi) & = &{1 \over (2 \pi R^2)^{3\over 2} }\, \exp(-\bxi^2/(2
R^2) ), \cr
\rho_p(\bpi) & = &{1 \over (2 \pi m T)^{3\over 2} }\, \exp(-\bpi^2/(2 m
T) ).
\eea
These expressions are given in the frame where 
we have a  non-expanding
static source  at rest.  

A multiplicity distribution when
Bose-Einstein effects are switched {off} (denoted by $p_n^{(0)}$), 
is a {FREE choice} in the model. We assume independent emission,
\be
        {p}^{(0)}_n \, = \,{n_0 ^n \over n!} \exp(-n_0).
\ee
 This completes the specification of the model.

\section{ Solution of Recurrences: } 

The probability of finding events with multiplicity $n$,
as well as the single-particle and the two-particle momentum distribution in
such  events is given as 
\bea
	N^{(n)}_1(\bk)  &  = &  
                \sum_{i=1}^n  {\dst  p_{n-i} \ov p_{n}}
		 G_i (\bk,\bk) , \label{e:d.2} \\
	N^{(n)}_2({\bk}_1,{\bk}_2)   &  =  & 
		\sum_{l=2}^n 
		\sum_{m=1}^{l-1} 
		{\dst p_{n-l}  \ov p_n}
		\left[ 
			G_m({\bk}_1,{\bk}_1) G_{l-m} ({\bk}_2,{\bk}_2) 
                + \right. \nonumber \\
	\null & \null &  
		\phantom{\sum_{l=2}^n 
		\sum_{m=1}^{l-1} 
		{\dst \omega_{n-l}  \ov \omega_n}}
		\left.
		G_m({\bk}_1,{\bk}_2) G_{l-m}({\bk}_2,{\bk}_1)\right] 
		. \label{e:d.3}
\eea

In ref.\cite{Pratt} recurrence relations were given 
for the construction of the functions $  G_m({\bk}_1,{\bk}_2)$.
In refs.\cite{Zim}, \cite{Tam} an explicit analytic form was 
obtained for these functions. We present now this method.

To arrive to this solution we
introduce the following auxiliary quantities:
 \bea
	\gamma_{\pm} 
		 \, = \, {\dst 1\ov 2} \left( 1 + x \pm \sqrt{1 + 2 x} \right),
			\label{e:gam.s} & \qquad  & 
	x \, = \,  R_{p}^2 \sigma_T^2 .
			\label{e:di.x}
 \eea
Using the notation
\bea
\sigma_T \, = \, 2  m T_p, 
\qquad	T_{p} \, =\, T + {\dst \sigma^2 \ov 2 m },
	& \qquad & 
        R_{p}^2 \, = \,
	 R^2 + {\dst m T \ov \sigma^2 \sigma_T^2},  \label{e:reff}
\eea
we arrive to the formulae of the {\it p}lane-wave 
pion laser model of Pratt, ref.~\cite{Pratt},
 if we replace $R $ in ref.~\cite{Pratt} with $ R_{p}$ and 
$ T $ in ref.~\cite{Pratt} with $ T_{p} $.  
	The {\it general analytical solution} of the model is given through
	the generating function of $p_n$, 
 \bea
	G(z) & = & \sum_{n = 0}^{\infty} p_n z^n  \label{e:d.g}
	\, = \, \exp\left( \sum_{n=1}^{\infty} C_n (z^n - 1) \right),
	\label{e:gsolu}
 \eea
	where $C_n$ is the combinant~\cite{GyK,Hegyi} of order $n$, 
 \be
	 C_n \, = \, {\dst n_0^n \ov n} 
			\left[ \gamma_+^{n\over 2} - 
				\gamma_-^{n\over 2} \right]^{-3 },
		\label{e:c.s} 
 \ee
	and the {\it general analytic solution} 
	for the functions $G_n({\bk}_1,{\bk}_2)$ reads as:
 \bea
	G_n({\bk}_1,{\bk}_2) & = &  j_n 
	\mbox{\rm e}^{ 
	- {b_n \over 2} \left[ 
		\left(\gamma_+^{n\over 2} \bk_1 
		- \gamma_-^{n \over 2} \bk_2\right)^2
		+ \left(\gamma_+^{n\over 2} \bk_2 - 
		  \gamma_-^{n \over 2} \bk_1\right)^2 \right]
	}, \\
	j_n  & = &  n_0^n \left[{ b_n \over \pi}\right]^{3 \over 2} ,
	\qquad 
	b_n \,  = \, {\dst 1 \over  \sigma_T^2} 
			{\dst \gamma_+ - \gamma_- \ov \gamma_+^n - \gamma_-^n}.
 \eea
The detailed proof that 
 the analytic solution to the  multi-particle wave-packet  model is indeed
given by the above equations is described in refs.~\cite{Zim}, \cite{Tam}. 

With our analytic solution we decreased 
the algorithmic complexity of the problem of $ n $ symmetrized boson system
from the original $ n!$ to $ 1 $.

\section{Multiplicity Distribution: 
}

	The mean and the second factorial moments are defined as follows 
\bea
	\langle n \rangle & = & \sum_n n\, p_n 
		\, = \, \sum_{i=1}^{\infty} i C_i, \\
	\langle n (n-1) \rangle & = & 
		\sum_n n\, (n-1)\, p_n \, = \, 
		 \langle n \rangle^2 +
		 \sum_{i=2}^{\infty} i\, (i-1)\, C_i. 
\eea
	The {large $n$ behavior} of the multiplicity 
	distribution,  $p_n $,  depends on 
	{$ n_0 / \gamma_+^{3\over 2}$}. This multiplicity distribution
has an interesting property. To see this, we define the quantity
$ n_c $ as  
 \bea
	n_c & = & \gamma_+^{3\over 2} = \left[ {\dst 1 + x + \sqrt{1 + 2 x} 
			\ov 2 } \right]^{3\over 2}.
 \eea
One can show that $ \langle n \rangle $ is finite, if  $n_0 < n_c$ 
and  $ \langle n \rangle $ is infinite, if  $n_0 > n_c$   
Hence $n_c$ is a {critical value} for the parameter $n_0$. 

\section{Solution for the Inclusive Distributions:}

In the previous sections we obtained results 
the multiplicity distribution and for exclusive momentum
distributions. To obtain the inclusive distribution 
we  introduce the auxiliary quantity
\be
	G(\bk_1,\bk_2) \, = \, \sum_{i = 1}^{\infty} G_i(\bk_1,\bk_2).
	\label{e:d.ggg}
\ee
( Higher order
Bose-Einstein symmetrization effects are negligible, if the first term
dominates the above infinite sum, i.e. if $G(\bk_1,\bk_2) = 
G_1(\bk_1,\bk_2)$.)

Now the averaging over the exclusive distributions can be
performed, 
\bea
	N_1(\bk) & = &  \sum_{n=1}^{\infty} p_n N_1^{(n)}(\bk) \, = \,
		  G(\bk,\bk), 
	\label{e:d.i.n1}
\eea
and the two-particle inclusive 
correlation functions can be evaluated 
as
\bea
	C_2(\bk_1,\bk_2)  & = &  
		{\dst N_2(\bk_1,\bk_2) \ov N_1(\bk_1) \, N_1(\bk_2) } 
	\nonumber \\
	\null & = & 
		1 +
		{ \dst G(\bk_1,\bk_2) G(\bk_2,\bk_1) \ov
                G(\bk_1,\bk_1) G(\bk_2,\bk_2) } .
		\label{e:d.i.c}
\eea 
This is an {exact result}, obtained 
{without the random phase} approximation
and without assuming that the number of sources is large.
However, this result is valid only when Bose-Einstein
condensation does not play a role, i.e. when 
$n_0 < n_c = \gamma_+^{3/2}$.

\section{ Rare gas limiting case:}

 For { large source sizes} or 
 { large effective temperatures},
 $ x >> 1$ we have  :
 \bea
	G_n(\bk_1,\bk_2) & \propto & 
		\left({\dst 2 \ov x}\right)^{{3\over 2}(n-1)} 
		\exp\left[ 
		 - {\dst n \ov 2 \sigma_T^2 }( \bk_1^2 + \bk_2^2) 
		- {R_{p}^2 \ov 2 n} 
			\left(\bk_1 - \bk_2\right)^2 \right],
	\label{e:g.i.rare} \\
	C_n  & = &  { \dst n_0^n  \ov n^4 } 
		\left({\dst 2 \ov x} \right)^{{3\over 2}(1-n)}.
	\label{e:c.rare}
 \eea
From this equation we can see that the effective temperatures
 and the  effective radii are decreased by a factor of $1/n$ for
$n$-th order symmetrization effects.
	
	The multiplicity distribution is shifted to high multiplicities: 
\be
	p_n  =  {\dst n_0^n \ov n!} \exp(-n_0) \,
		\left[ 1 + {\dst n(n-1) - n_0^2 \ov 2 (2 x)^{3\over 2}
			} \right].
	\label{e:pn.sol}
\ee

The momentum distribution is  obtained in this approximation
by an expansion in the small parameter
	$ \eps = \left( 2 /  x  \right)^{3/2}$. 
In this approximation we get 
\bea
	N_1(\bk) & = & 
		{\dst n_0 \ov ( \pi \sigma_T^2)^{3\over 2} }
		\exp\left( -{\dst  \bk^2  \ov \sigma_T^2} \right) +
		{\dst n_0^2 \ov (\pi \sigma_T^2 x )^{3\over 2} }
		\, \exp\left( - {\dst 2 \ov \sigma_t^2} \bk^2 \right)
		\label{e:n.i.rare}\\
	N_1^{(n)}(\bk)  & = & 
	{\dst n \ov ( \pi \sigma_T^2)^{3 \over 2} }
			\exp\left( -{\dst  \bk^2  \ov \sigma_T^2} \right) 
		\left\{ 1  +  {\dst (n-1)\ov (2 x)^{3 \over 2} } 
		 \left[ 
		2^{3 \over 2}
		\exp\left( -{\dst \bk^2 \ov \sigma_t^2}\right) 
		 -1 \right] \right\}. 
\eea
	
The single-particle inclusive and exclusive
	 momentum distributions are {enhanced at low momentum.} 
The influence of the wave packet size on
the  pion multiplicity distribution is shown in Fig.1.
The effect of the wave packet width on the momentum 
distribution is displayed on Fig.2.

\begin{figure}
\epsfig{file=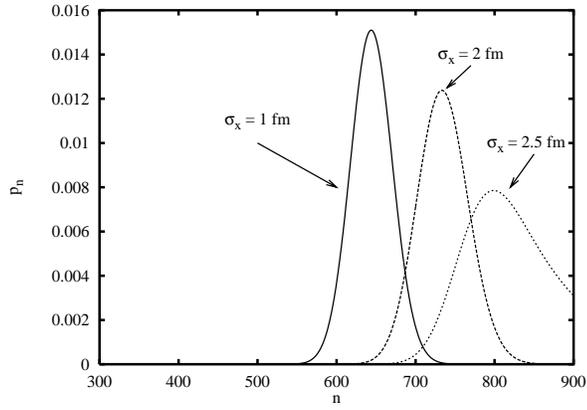,height=5.6cm,width=8.cm}
\caption{
As the size of the wave packets is changed, the overlap changes and 
stimulated emission results in larger multiplicities.
}
\label{fig:1}
\end{figure}

\begin{center}
\begin{figure}
\epsfig{file=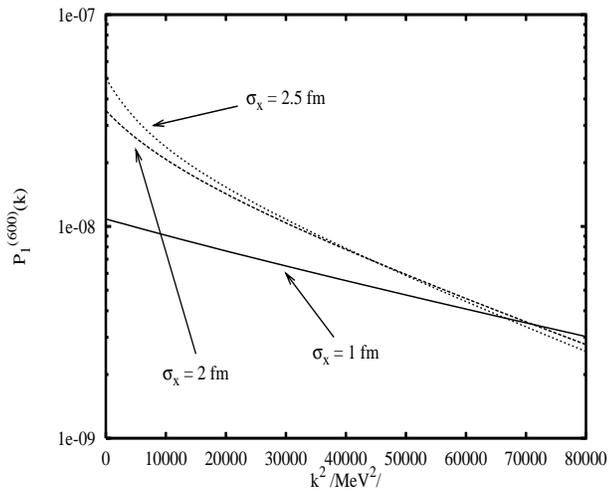,height=6.7cm,width=8.1cm }
\caption{
Stimulated emission results in enhancement of pions in the low momentum
modes.}
\label{fig:2}
\end{figure}
\end{center}

\section{Correlation functions: }

In the highly condensed  $ x << 1 $  and $n_0 >> n_c$ 
Bose gas limit a kind of lasing behaviour  and an optically coherent behaviour
is obtained, which is characterized by \cite{Tam}
the disappearance of the bump in the correlation function:
{
	$$C_2(\bk_1,\bk_2) = 1.$$}

On the other hand, in the {rare gas} limit we get
\bea 
	C_2(\bk_1,\bk_2)  & = &  
	1  +   \exp\left( - { R_{p}^2 \dek^2 
		} \right) .
	\label{e:c.li}
\eea

\subsection{Exclusive Correlation Functions: }

In a highly condensed limiting case, $ x << 1 $ and $n_0 >> n_c$, the 
{exclusive and inclusive}
correlation functions {coincide}, $ C_2(\bk_1,\bk_2) = 1$.
 
 In the { rare gas} limit, $ x >> 1$, the 
{exclusive} and {inclusive} correlations
are {different}, and the exclusive correlation has the form
\bea
C^{(n)}_2(\bk_1,\bk_2)  & = &  {\dst n^2 \ov n(n-1) } 
		{\dst N_2^{(n)}(\bk_1,\bk_2) \ov
		N_1^{(n)}(\bk_1) \, N_1^{(n)}(\bk_2) } \nonumber \\
 \null & = &  1 + \lambda_{\bak} 
		\exp\left( - R_{\bak,s}^2  \dek_{s}^2 
		- R_{\bak,o}^2 \dek_{o}^2 \right), 
\eea
where $\bak = 0.5 (\bk_1 + \bk_2)$, 
{$\dek_{s} = \dek - \bak{ (\dek \cdot \bak) / ( \bak \cdot \bak)} $} and 
{$\dek_{o} = \bak{ (\dek \cdot \bak) / ( \bak \cdot \bak)} $},

The {momentum dependent} parameters are  given as follows:
\bea
\lambda_{\bf{K}}  & = &   1 +
		{\dst 2 \ov ( 2 x)^{3\over 2} }
		\left[ 1 - 
		2^{(5/2)} \exp\left( - {\dst {\bf K}^2 \ov \sigma_T^2 } \right)
		\right], \label{e:lambda.corr} \\
R_{{\bf{K}},s}^2  & =   &  R_{p}^2   + 
		{\dst 1 \ov ( 2 x)^{3\over 2} }
	 	\left[  R_{p}^2  -  \sqrt{2} 
		\exp\!\left( - {{\bf K}^2 \ov \sigma_T^2} \right)	 \!
		\left(  
		 ( n + 2) R_{p}^2  +  {\dst 2 \ov \sigma_T^2 }
		\right) 
		\right], \label{e:r.corr}\\
R_{{\bf K},o}^2  & = &  R_{{\bf{K}},s}^2 +
		{\dst n \ov    x^{3\over 2} } {\dst {\bf K}^2  \ov \sigma_T^4 } 
		\exp\left( - {\dst {\bf K}^2 \ov \sigma_T^2} \right) .
		\label{e:c.dir}
\eea

The dependence of the parameters of the correlation functions
on the mean  momentum of the two pions, ${\bf{K}}$, is shown on Fig.2. 
The influence of the wave packet size on the two particle 
correlation is illustrated on Fig.3. 
(These limiting cases are discussed in more details in ref.\cite{Zim}.)

\begin{center}
\begin{figure}[t]
\epsfig{file=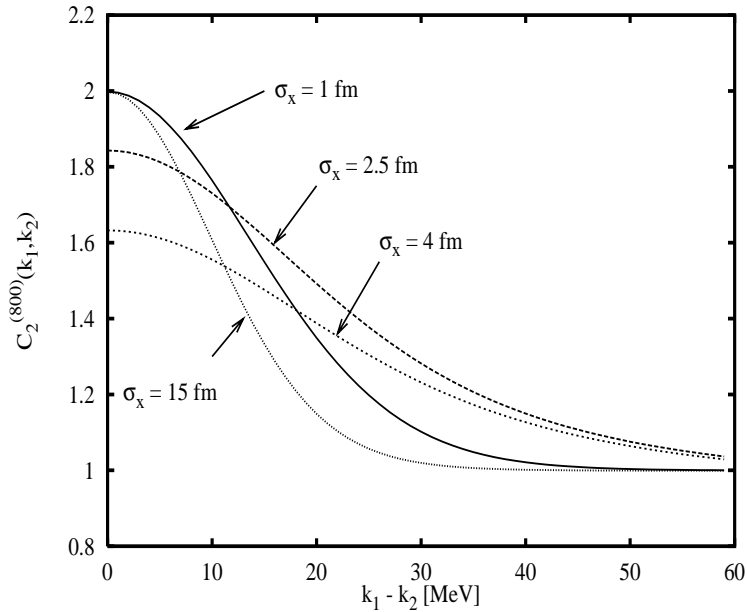,height=8.4cm,width=10.15cm }
\caption{
 $C^{(800)}_2(\bk_1,\bk_2)$ 
 for wave packet sizes $ \sigma_x = 1.0$, $2.5$, $4.0 $  and $ 15.0 $ fm
 (solid, dashed, dotted and dense-dotted lines).
$ | \bk_1 + \bk_2 | / 2 = 50$ MeV,  out component.
 The number of pions was set to $n=800$. 
}
\label{fig:3}
\end{figure}
\end{center}

\begin{center}
\begin{figure}[t]
\epsfig{file=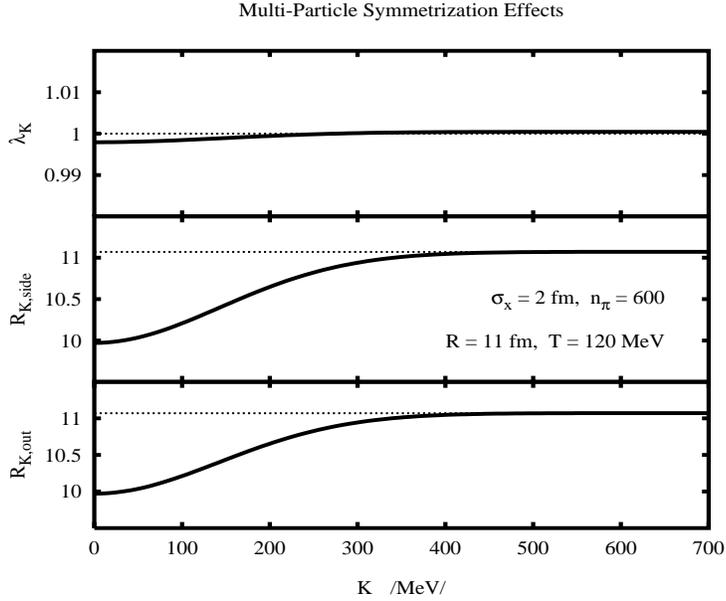,height=8.4cm,width=10.15cm }
\caption{
	Momentum-dependent
	reduction of the intercept parameter $\lambda_{\bak}$,
	the side-wards and the outwards
	radius parameters, $R_{\bak,s}$ and $R_{\bak,o}$
	from their static values of 1 and $R_{p}$, respectively.
}
\label{fig:4}
\end{figure}
\end{center}

\section{Highlights: }

In this paper we presented a consequent quantum mechanical 
description of the correlations caused by the multi-particle
Bose-Einstein symmetrization
of a system of large number of bosons.

We introduced the overlapping  multi-particle wave-packet 
density matrix describing stimulated emission. 
We reduced the algorithmic complexity of
 the description of the $ n $ boson states from $ n! $ to $ 1 $ by 
obtaining an explicite analytical solution for the problem.  

 We have shown, that the
radius and intercept parameters  depend on the mean momentum
of the two pions even for static sources, due to multi-particle
symmetrization effects.

We have found an enhancement 
 of the wave-packets in the low momentum modes,
due to  multi-particle Bose-Einstein symmetrization.
When all pions are in the lowest momentum state, the 
system may be interpreted as a pion-laser.

\section*{Acknowledgments}

The present work was supported in part by 
the US - Hungarian  Joint Fund MAKA 652/1998, 
by the National Scientific Research Fund (OTKA, Hungary) Grant 024094,
and by NWO (Netherland) - OTKA grant N25487.
The authors thank these sponsoring organizations for their aid.

\end{document}